\documentstyle[prl,aps,psfig]{revtex}
\begin{document}
\twocolumn[\hsize\textwidth\columnwidth\hsize\csname
@twocolumnfalse\endcsname
\title{Self-organized critical neural networks}
\author{Stefan Bornholdt$^{1,2,a}$
and Torsten R\"ohl$^1$}
\address{$^{1}$Institute for Theoretical Physics, University of Kiel,
Leibnizstrasse 15, D-24098 Kiel, Germany \\
$^{2}$Interdisciplinary Center for Bioinformatics, University of 
Leipzig,
Kreuzstrasse 7b, D-04103 Leipzig, Germany\\
$^{a}$Email address: bornholdt@izbi.uni-leipzig.de}
\maketitle
\begin{abstract}
A mechanism for self-organization of the degree of connectivity in
model neural networks is studied. Network connectivity is regulated
locally on the basis of an order parameter of the global dynamics which
is estimated from an observable at the single synapse level.
This principle is studied in a two-dimensional neural network 
with randomly wired asymmetric weights. In this class of networks,  
network connectivity is closely related to
a phase transition between ordered and disordered dynamics.
A slow topology change is imposed on the network through a local
rewiring rule motivated by activity-dependent synaptic development:
Neighbor neurons whose activity is correlated, on average develop 
a new connection while uncorrelated neighbors tend to disconnect.
As a result, robust self-organization of the network towards the 
order disorder transition occurs. Convergence is independent of 
initial conditions, robust against thermal noise, and does not 
require fine tuning of parameters.
\medskip \\
Published in: Phys.\ Rev.\ E 67 (2003) 066118.  
\end{abstract}
\medskip
]

Information processing in living organisms is often performed by
large networks of interacting cells with an overall stunning degree of
complexity. How can such networks be efficiently constructed and
how can a robust functioning be ensured?
The observed complexity of many nervous systems exceeds by far
what can be hard coded in the genome \cite{numberofgenes}.
Therefore, developmental principles play a key role in
network construction. Furthermore, as learning is a major function
of such networks, self-organization and adaptation processes
continue throughout the lifetime of a network.

But how can robustness of large dynamical networks be
ensured in the face of continuous developmental and
adaptive processes? In general, dynamical
stability of large networks of dynamical elements and 
robustness against perturbations are not obtained for free:
Model networks with asymmetric connectivity patterns often
exhibit regimes of chaotic dynamics with large parameter ranges
where network dynamics is not easily controlled \cite{nnnoise}.
In networks whose central function is information transfer, 
these regimes would instantly render them useless.
Consider, for example, model neural networks with asymmetric
synaptic couplings, where a percolation transition between regimes
of ordered and disordered dynamics is known \cite{kuerten}.
In the disordered phase, which occurs for densely connected networks,
already small perturbations percolate through
the networks.\footnote{This is reminiscent of avalanche like
propagation of activity in the brain which is observed in some
diseases of the central nervous system \cite{epilepsy}.}
In such networks, developmental processes that change connectivity
always face the risk of driving the network into the highly connected 
regime (where chaotic dynamics prevails), as long as no explicit 
mechanism is given that controls the global degree of connectivity.

We here study this question of dynamical robustness of
networks in the presence of developmental processes
in the context of a simple toy model,
an asymmetric neural network combined with simple 
topology-changing rules. In particular we ask how a local rewiring
mechanism could control global dynamical properties of a large
network and actively contribute to avoiding chaotic regimes.
While an obvious possibility is a direct feedback of the global 
dynamical
state to the synapses, e.g., controlling synaptic growth rates,
we here consider an
even simpler mechanism that relies on local information only
and in principle could be at work in natural systems. We argue
that if an order parameter characterizing a global phase transition is
accessible at the single synapse level, it can provide the basis for
a regulation of global network connectivity solely on the basis of
local mechanisms.

Recent models of self-organization of network structures show that
it is possible to locally measure a global order parameter connected
to the percolation transition of the network, namely the average 
activity of a single node over time \cite{bornholdt00}.
Here we will see that, similarly, the average correlation between 
the activities of two neurons contains information about the 
global order parameter as well.
The network can then use this approximate order parameter
to guide the developmental rule. An interesting question is 
whether self-organization to a critical
dynamical transition could occur in a model neural network on the
basis of such a correlation. A possible rule is that
new synaptic connections preferentially grow between
correlated neurons,  as suggested by
the early ideas of Hebb \cite{hebb49} and the observation of
activity-dependent neural development \cite{development}.  
In the remainder of this paper let us study this problem in the
framework of a specific toy model. We will first define a neural network
model with a simple mechanism of synaptic development.
Then, with numerical studies we will discuss the interplay of
dynamics on the network with dynamics of the network topology.
Finally, robustness of self-organizing processes in this model
and possible implications for biological systems are discussed.

Let us consider a two-dimensional neural network with random
asymmetric weights on the lattice. The neighborhood of each neuron
is chosen as its Moore neighborhood with eight neighbors.\footnote{The
choice of the type of neighborhood is not critical,
however, here the Moore neighborhood is more convenient than the
von Neumann type since, in the latter case, the critical link
density (fraction of nonzero weights) at the percolation threshold
accidentally coincides with the attractor of the trivial developmental 
rule of producing a link with $p=0.5$.  In general, also random sparse
neighborhoods would work as demonstrated in Ref.\ \cite{bornholdt00}.} 
The weights $w_{ij}$ are randomly drawn from a uniform
distribution $w_{ij} \in \left[-1, +1 \right]$ and are nonzero
between neighbors, only. Note that weights $w_{ij}$ are 
asymmetric, i.e., in general,  $w_{ij} \not= w_{ji}$.  
Within the neighborhood of a node, a fraction of its weights 
$w_{ij}$ may be set to $0$. The average number of nonzero 
weights per node is called the average connectivity $K$ of the network 
(for definiteness count e.g.\ the incoming weights at each node, only).  
The network consists of $N$ neurons with
states $\sigma_i = \pm 1$ which are updated in parallel with a
stochastic Little dynamics on the basis of inputs received from the 
neighbor neurons at the previous time step:
\begin{eqnarray}
\mbox{prob}[\sigma_i(t+1)=+1] &=& {g_\beta}\left(f_i(t)\right)
\nonumber \\
\mbox{prob}[\sigma_i(t+1)=-1] &=& 1-{g_\beta}\left(f_i(t)\right)
\end{eqnarray}
with
\begin{eqnarray}
f_i(t) = \sum_{j=1}^N w_{ij}\sigma_j(t) + \theta_i
\end{eqnarray}
and
\begin{eqnarray}
g_\beta( f_i(t)) = \frac{1}{1 +  e^{- 2 \beta f_i(t)}}
\end{eqnarray}
with the inverse temperature $\beta$ and a threshold $\theta_i$.
The threshold is chosen here as $\theta_i= -0.1+\gamma$
and includes a small random noise term $\gamma$ from
a Gaussian of width $\epsilon$. This noise term is motivated by
the slow fluctuations observed in biological neural
systems \cite{plasticity}. With respect to varying either 
$\theta$ or $K$, the network exhibits a percolation
transition between a phase of ordered dynamics, with short transients
and short limit cycle attractors, and a phase of chaotic dynamics where
the length of dynamical patterns scales exponentially with system size
\cite{kuerten,thresholdnets}.

\begin{figure}[htb]
\let\picnaturalsize=N
\def\picsize{80mm}
\def\picfilename{fig3.eps}
\ifx\nopictures Y\else{\ifx\epsfloaded Y\else\input epsf \fi
\let\epsfloaded=Y
\centerline{\ifx\picnaturalsize N\epsfxsize \picsize\fi
\epsfbox{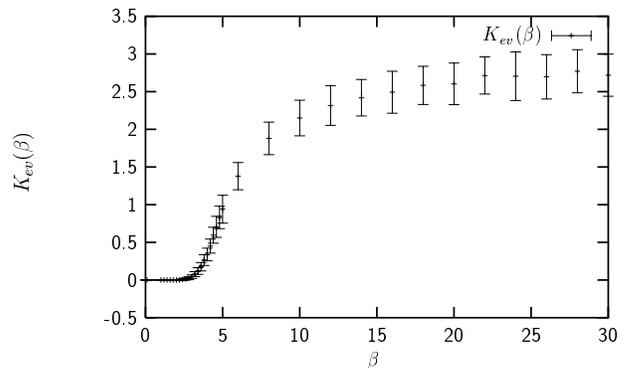}}}\fi
\caption{Evolved average connectivity $K_{ev}$ as a function of
the inverse temperature $\beta$. Each point is averaged over
$10^5$ time steps in a network of size $N=64$ and $\alpha=0.5$.}
\label{temperature}
\end{figure}

The second part of the model is a slow change of the topology of the
network by local rewiring of synaptic weights: If the activity of
two neighbor neurons is on average highly correlated (or 
anticorrelated),
they will obtain a common link. If their activity on average is less 
correlated, they will lose their common link. To be more specific,
let us define the average correlation $C_{ij}(\tau)$ of a pair $(i,j)$
of neurons over a time interval $\tau$
\begin{eqnarray}
C_{ij}(\tau) = \frac{1}{\tau + 1 }\sum_{t=t_0}^{t_0+\tau
}\sigma_i(t)\sigma_j(t).
\end{eqnarray}
The full model dynamics is then defined as follows. 
\begin{enumerate}
\item
Start with a random network with an average connectivity
(number of nonzero weights per neuron) $K_{ini}$
and a random initial state vector $\vec{\sigma}(0) =
(\sigma_1(0),...,\sigma_N(0))$.
\item
For each neuron $i$, choose a random threshold $\theta_i$ from a
Gaussian distribution of width $\epsilon$ and mean $\mu$.
\item
Starting from the initial state, calculate the new system state
applying eq.\ (1) using parallel update. Iterate this for
$\tau$ time steps.
\item
Randomly choose one neuron $i$ and one of its neighbors $j$ and
determine the average correlation $C_{ij}(\tau/2)$ over the last
$\tau/2$ time steps. (Alternatively, the correlation can be obtained
from a synaptic variable providing a moving average at any given time).
\item
If $|C_{ij}(\tau)|$ is larger than a given threshold $\alpha$,
$i$ receives a new link $w_{ij}$ from site $j$ with a weight 
chosen randomly from
the interval $w_{ij} \in \left[-1,1\right]$.\footnote{Also binary 
weights could
be used as in Ref.\ \cite{bornholdt00}.}
If $|C_{ij}(\tau)|\leq \alpha$, the link $w_{ij}$ is set to $0$ 
(if nonzero).  
\item
Go to step 2 and iterate, using the current state of the network
as new initial state.
\end{enumerate}
\noindent
The dynamics of this network is continuous in time, with 
neuron update on a fast time scale and topology update 
of the weights on a well-separated slow ``synaptic 
plasticity'' time scale. Note that the topology-changing rule does 
not involve any global knowledge, e.g., about attractors.
A typical scenario of this dynamical evolution is shown in Fig.\ 1
\begin{figure}[t]
\let\picnaturalsize=N
\def\picsize{85mm}
\def\picfilename{fig1.eps}
\ifx\nopictures Y\else{\ifx\epsfloaded Y\else\input epsf \fi
\let\epsfloaded=Y \centerline{\ifx\picnaturalsize N\epsfxsize
\picsize\fi \epsfbox{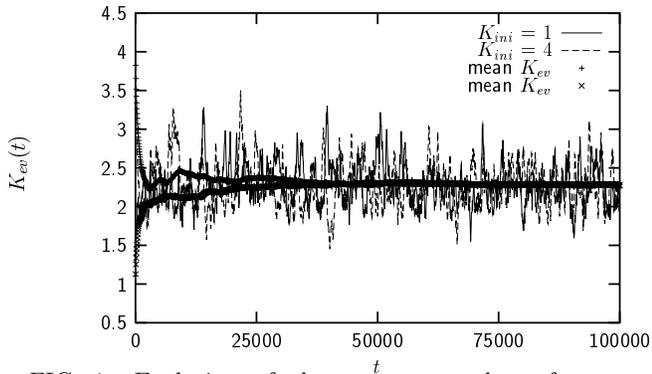}}}\fi
\caption{Evolution of the average number of nonzero weights per
neuron over evolutionary time, for a system of size $N=64$ $(8\times8)$
and two different initial connectivities ($K_{ini}=1.0$ and 
$K_{ini}=4.0$).
Independent of the
initial conditions the networks evolve to a specific average
connectivity. Parameters are $\beta = 25$, $\epsilon=0.1$, a correlation
cutoff $\alpha = 0.8$, and an averaging time window of $\tau=200$.}
\end{figure}
where the average number of nonzero weights per neuron $K_{ev}$ is
shown as a time series and as cumulative mean.
One observes that the continuous network dynamics, including the
slow local change of the topology, results in a convergence of the
average connectivity of the network to a characteristic value which is
independent of initial conditions.
\begin{figure}[htb]
\let\picnaturalsize=N
\def\picsize{85mm}
\def\picfilename{fig2.eps}
\ifx\nopictures Y\else{\ifx\epsfloaded Y\else\input epsf \fi
\let\epsfloaded=Y
\centerline{\ifx\picnaturalsize N\epsfxsize \picsize\fi
\epsfbox{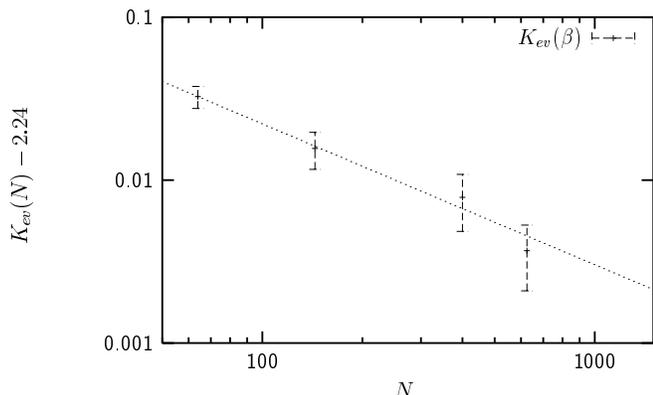}}}\fi \caption{Finite size scaling of the evolved
average connectivity. Averages are taken over $4\times 10^5$ time steps.}
\label{kconvergence}
\end{figure}
Finite size scaling of the resulting average connectivity indicates
the convergence towards a characteristic value for large network size 
$N$
and exhibits the scaling relationship
\begin{eqnarray}
K_{ev}(N) = a N^{-\delta} + b
\end{eqnarray}
with $a =1.2\pm0.4$, $\delta =0.86\pm0.07$, and $b=2.24\pm0.03$.
Thus, in the large system size limit $N \rightarrow \infty$ the
networks evolve towards $K_{ev}^\infty = 2.24\pm0.03$
(see Fig.\ \ref{kconvergence}).
The self-organization towards a specific average connectivity is
largely insensitive to thermal noise of the network dynamics, up to
$\approx 10 \%$ of thermal switching errors (or $\beta >10$)
of the neurons.
This indicates that the structure of a given dynamical attractor 
is robust against a large degree of noise.
Figure \ref{temperature} shows the evolved average
connectivity as a function of the inverse temperature $\beta$.
While the stability of dynamical attractors on an intermediate time
scale is an important requirement for the local sampling of neural
correlation, on the long time scale of global topological changes,
switching between attractors is necessary to ensure ergodicity at 
the attractor sampling level.  The second source of noise, the slow
random change in neural thresholds as defined in step (2) of the
algorithm, is closely related to such transitions between attractors.
While, in general,
the model converges also when choosing some arbitrary fixed threshold
$\theta$ and omitting step (2) from the algorithm,
a small threshold noise facilitates
transitions between limit cycle attractors \cite{thresholdnoise}
and thus improves sampling over all attractors of a network,
resulting in an overall increased speed and robustness of the 
convergence.
An asynchronous change of the threshold
$\theta_i$, updating one random $\theta_i$ after completing one
sweep (time step) of the network, leads to similar results
as the parallel rule defined above.

\begin{figure}[htb]
\let\picnaturalsize=N
\def\picsize{85mm}
\def\picfilename{fig4.eps}
\ifx\nopictures Y\else{\ifx\epsfloaded Y\else\input epsf \fi
\let\epsfloaded=Y
\centerline{\ifx\picnaturalsize N\epsfxsize \picsize\fi
\epsfbox{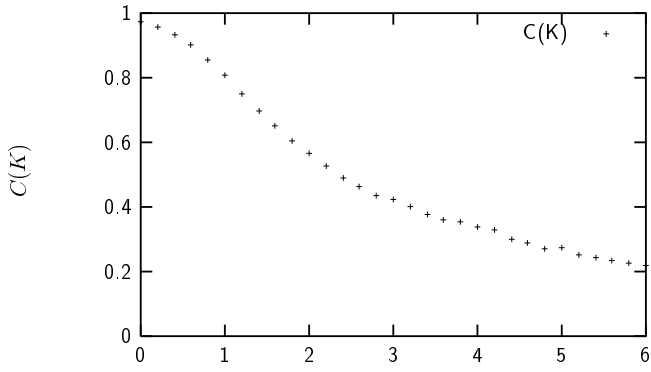}}}\fi
\caption{The average correlation $|C_{ij}(\tau)|$ between random neurons
of random networks at different connectivities $K$. Samples are taken
over 1000 random networks with 100 random initial conditions each,
for network size $N=64$.}
\label{correlation}
\end{figure}

\begin{figure}[htb]
\let\picnaturalsize=N
\def\picsize{85mm}
\def\picfilename{fig5.eps}
\ifx\nopictures Y\else{\ifx\epsfloaded Y\else\input epsf \fi
\let\epsfloaded=Y
\centerline{\ifx\picnaturalsize N\epsfxsize \picsize\fi
\epsfbox{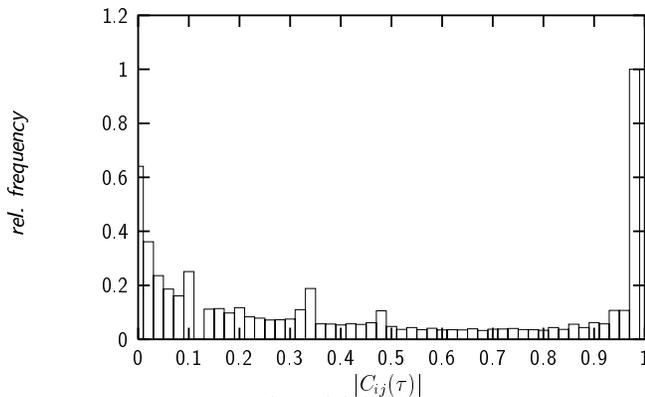}}}\fi
\caption{Histogram of $\mid C_{ij}(\tau)\mid $ for a network
evolving in time, with
$N=64$ and $\beta=10$, taken over a run of $4\times10^5$ time
steps.}
\label{alpha}
\end{figure}

The basic mechanism of the observed self-organization in this system is
the weak coupling of topological change to an order parameter of the
global dynamical state of the network, and thus is different from the
mechanism of extremal dynamics, underlying many prominent
models of self-organized criticality \cite{soc}. To illustrate this,
let us for a moment consider the absolute average correlation
$|C_{ij}(\tau)|$ of two neurons which is the parameter used as a
criterion for the rewiring process. For random networks, this quantity 
is shown in Fig.\ \ref{correlation} for different connectivities $K$.
Note that the correlation is large for networks with small connectivity,
and small for networks that are densely connected.
The rewiring rule balances between these two regimes: 
For high correlation, it is more likely that a link is created, 
at low correlation, links are vanishing. 
The balance is reached most likely in the region of the curve 
where the slope reaches its maximum, as here the observed correlation 
reacts most sensitively to connectivity changes. As the steep portion
of the correlation curve occurs in a region of small connectivities 
where also the critical 
connectivity $K_c\approx 2$ of the network is located, 
this makes the correlation measure sensitive to the global dynamical 
state of the network and potentially useful as an approximation of 
the order parameter. Synaptic development dependent on averaged 
correlation between neurons can thus obtain approximate information 
about the global 
dynamical state of the network as is realized in the 
above toy model with a simple implementation on the basis of
a threshold $\alpha$. The exact choice of the threshold $\alpha$ is
not critical, which can be seen from the histogram of the absolute
correlation $|C_{ij}(\tau)|$ shown in Fig.\ \ref{alpha} for a
typical run of the model.
Correlations appear to cluster near high and near low values such
that the cutoff can be placed anywhere inbetween the two regimes.
Even a threshold value close to $1$, as compared with the correlation
cutoff $\alpha=0.8$ used in the simulations here, only leads to a
minor shift in $K_{ev}$ and does not change the overall behavior.

\begin{figure}[htb]
\let\picnaturalsize=N
\def\picsize{85mm}
\def\picfilename{fig6.eps}
\ifx\nopictures Y\else{\ifx\epsfloaded Y\else\input epsf \fi
\let\epsfloaded=Y \centerline{\ifx\picnaturalsize N\epsfxsize
\picsize\fi \epsfbox{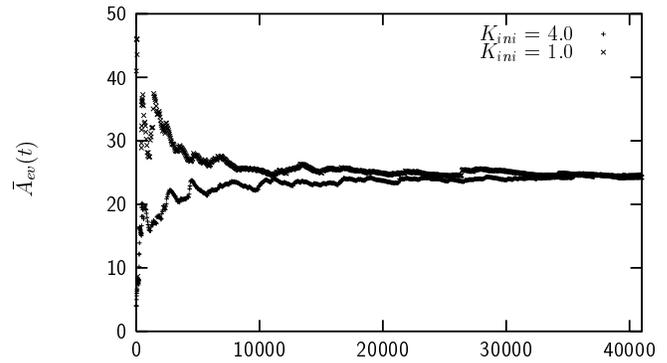}}}\fi
\caption{Evolution of the cumulative average of attractor length for 
the same
system as shown in Fig.\ 1. The mean attractor length converges to a
value independent of the two initial conditions of the network shown 
here.
The attractor
length is measured at zero temperature in order to have an exactly
defined measure.}
\label{attractors}
\end{figure}

\begin{figure}[htb]
\let\picnaturalsize=N
\def\picsize{85mm}
\def\picfilename{fig7.eps}
\ifx\nopictures Y\else{\ifx\epsfloaded Y\else\input epsf \fi
\let\epsfloaded=Y
\centerline{\ifx\picnaturalsize N\epsfxsize \picsize\fi
\epsfbox{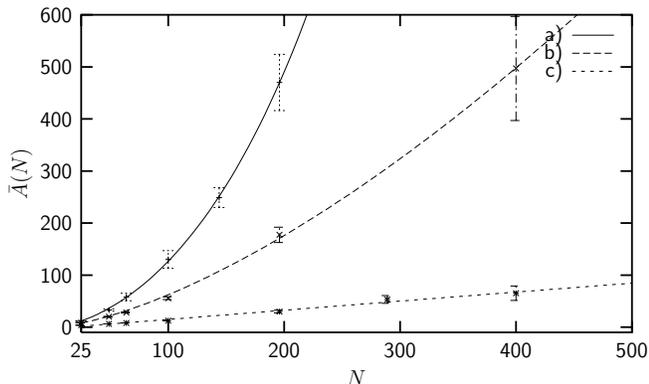}}}\fi
\caption{Finite size scaling of the evolved average attractor period
(b). Also shown for comparison is the corresponding scaling of the
attractor lengths of an overcritical random network (a) with $K=3.8$
and an undercritical one (c) with $K=1.5$. Symbols denote measured
values and lines correspond to the fits
$f_a(x)=15.1 x^{0.57} \left( e^{0.005 x}-1 \right)$,
$f_b(x)=0.6 x^{1.5}$, and
$f_c(x)=0.28 x^{0.75} \ln (0.097 x)$.
}

\label{phases}
\end{figure}

Up to now we focused on changes of the network structure as a result of
the dynamics on the network. A further aspect is how the structural
changes affect the dynamics on the network itself. Do also dynamical
observables of the networks self-organize as a result of the observed
convergence of the network structure? An interesting quantity in this
respect is the average length of periodic attractors as shown in Fig.\
\ref{attractors}.
Indeed, this dynamical observable of the network dynamics converges
to a specific value independent of the initial network, similarly to
the convergence of the structural parameter $K$ considered earlier.
From the $K$ dependency of the neural pair correlation we have seen
above that the rewiring criterion tends to favor connectivities near
the critical connectivity of the network.
Does also the evolved average attractor length relate to
critical properties of the percolation transition? An approximate
measure of this aspect is the finite size scaling of the evolved
average period as shown in Fig.\ \ref{phases}.
For static networks we find that
the attractor lengths typically scale exponentially with $N$
in the overcritical regime, but less than linearly
in the ordered regime. For the evolved connectivity $K_{ev}$
in our model, we observe scaling close to criticality.
Large evolved networks exhibit relatively short attractors,
which otherwise for random networks in the overcritical regime
could only be achieved by fine tuning.
The self-organizing model studied here evolves nonchaotic networks
without the need for parameter tuning.

To summarize, neural network development has been studied in
an asymmetric model neural network. The developmental rule is
based on local rewiring motivated by Hebbian, activity-dependent
synaptic development.
In a continuously running network, robust self-organization of the
network towards the percolation transition between ordered and
disordered dynamics is observed, independent of initial conditions
and robust against thermal noise. The basic model is robust against
changes in the details of the algorithm. We conclude that a weak
coupling of the rewiring process to an approximate measurement of
an order parameter of the global dynamics is sufficient for a robust
self-organization towards criticality.
In particular, the order parameter has been estimated
solely from information available on the single synapse level via time
averaging of correlated neural activities.

While here we considered self-organization in model neural networks,
the observed mechanism may occur in other more complex systems.
For example,
global dynamical order from self-organization at the synapse level
could, in principle, be at work in biological nervous systems as well.
Prerequisites are an averaging procedure
of correlated activities on slow time scales (similar to synaptic
processes underlying learning through long term potentiation),
and a coupling to synaptic development.

\end{document}